\documentclass[a4paper,preprint,aps,superscriptaddress]{revtex4}
\usepackage{epsfig}
\begin{document}

\title{Energies and radial distributions of $B_s$ mesons on the lattice}

\collaboration{UKQCD Collaboration}
\noaffiliation
\author{J. Koponen}
\email{jonna.koponen@helsinki.fi}
\affiliation{Department of Physical Sciences and
Helsinki Institute of Physics,
P.O. Box 64, FIN--00014 University of Helsinki,Finland}
\preprint{HIP-2007-03/TH}

\begin{abstract}
This is a follow-up to our earlier work for the energies and the charge (vector)
and matter (scalar) distributions for S-wave states in a heavy-light meson, where
the heavy quark is static and the light quark has a mass about that of the strange
quark. We now study excited states of these mesons with higher angular momentum
and with radial nodes.

The calculation is carried out with dynamical fermions on a $16^3 \times 32$
lattice with a lattice spacing $a\approx 0.11$~fm.
The lattice configurations were generated by the UKQCD Collaboration.
Attempts are now being made to understand these results in terms of the Dirac
equation.
In nature the closest equivalent of this heavy-light system is the $B_s$ meson,
which allows us to compare our lattice calculations to experimental results
(where available) or
give a prediction where the P-wave states should
lie. We pay particular attention to the spin-orbit splitting, to see which
one of the states (for a given angular momentum L) has the lower energy.
\end{abstract}
%\PACS{14.40.Nd, 11.15.Ha, 12.38.Gc}
\maketitle

\section{Motivation}

There are several advantages in studying a heavy-light system on a lattice.
Our meson is much more simple than in true QCD: one of the quarks is static with
the light quark ``orbiting'' it. This makes it very beneficial for modelling.
On the lattice an abundance of data can be produced, and we know which state we are
measuring -- the physical states can be a mixture of two or more configurations, but
on the lattice this complication is avoided. However, our results on the
heavy-light system can still be compared to the $B_s$ meson experimental
results.

\section{Measurements and lattice parameters}

We have measured both angular and radial excitations of heavy-light mesons,
and not just their energies but also some radial distributions.
Since the heavy quark spin decouples from the game we may label the states as
$\mathrm{L}_{\pm}=\mathrm{L}\pm\frac{1}{2}$, where L is the angular momentum
and $\pm\frac{1}{2}$ is the spin of the light quark.

The measurements were done on a $16^3\!\times\! 32$ lattice.
We have two degenerate quark flavours with a mass that is close
to the strange quark mass. The lattice configurations were generated by the UKQCD
Collaboration. More details of the different lattices used in this study can be found in
Refs.~\cite{PRD69}, and references therein.
Two different levels of fuzzing (2 and 8 iterations of conventional fuzzing) were
used in the spatial directions to permit the extraction of the excited states.

\section{2-point correlation function}

\begin{figure}[b]
\centering
\includegraphics*[width=0.31\textwidth]{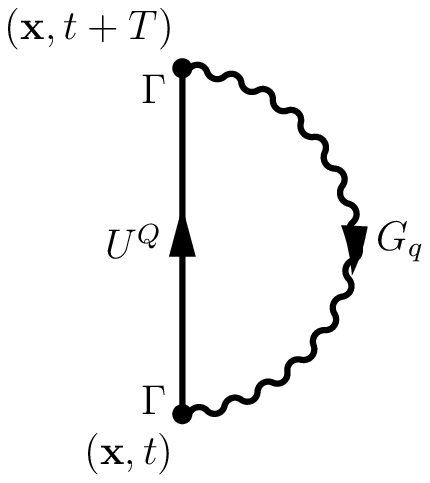}
\includegraphics*[width=0.31\textwidth]{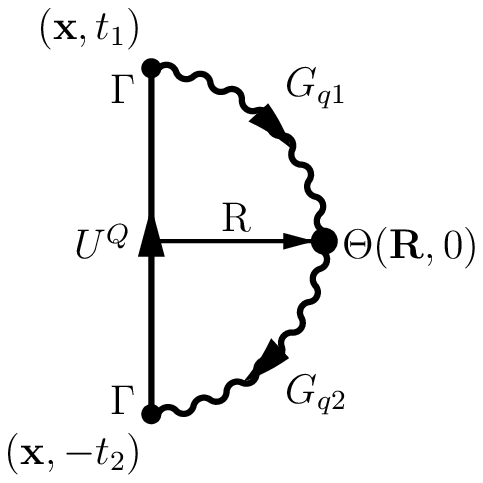}
\caption{On the left: Two-point correlation function.
On the right: Three-point correlation function.}
\label{Fig:Corr}
\end{figure}

The 2-point correlation function (see Figure~\ref{Fig:Corr}) is defined as
\begin{equation}
\label{2point}
C_2(T)=\langle P_t\Gamma G_q(\mathbf{x},t+T,t)P_{t+T}
\Gamma^{\dag}U^Q(\mathbf{x},t,t+T)\rangle \  ,
\end{equation}
where $U^Q(\mathbf{x},t,t+T)$ is the heavy (infinite mass)-quark propagator
and $G_q(\mathbf{x},t+T,t)$ the light anti-quark propagator. $P_t$
is a linear combination of products of gauge links at time $t$
along paths $P$ and $\Gamma$ defines the spin structure of the operator.
The $\langle ...\rangle$ means the average over the whole lattice.
The energies ($m_i$) and amplitudes ($a_i$) are extracted by fitting the $C_2$
with a sum of exponentials,
\begin{equation}
\label{C2fit}
C_2(T)\approx\sum_{i=1}^{N_{\textrm{max}}}a_{i}\mathrm{e}^{-m_i T},\;
\textrm{where $N_{\textrm{max}}=2\textrm{ -- }4$, $T\leq 14$}.
\end{equation}
Fuzzing indices have been omitted for clarity.

\section{Energy spectrum and spin-orbit splittings}

\begin{figure}
\centering
\includegraphics[height=0.55\textwidth, angle=-90]{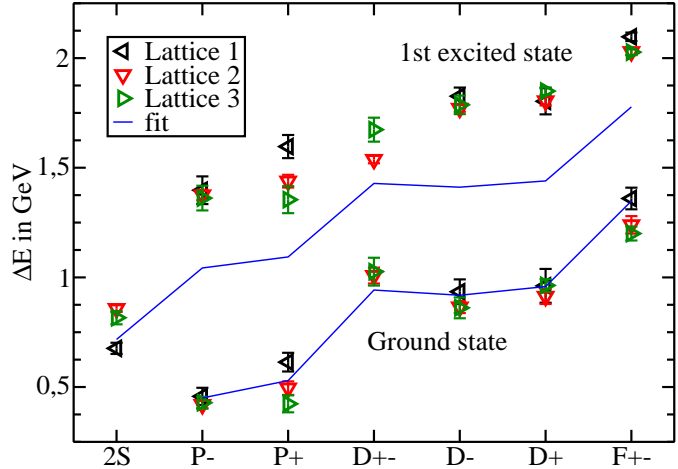}
\caption{Energy spectrum of the heavy-light meson.
Here L$+$($-$) means that the light quark spin couples to angular
momentum L giving the total $j=\textrm{L}\pm 1/2$. 2S is the first radially
excited L$=0$ state. The D$+-$ is a mixture of the D$-$ and D$+$ states, and likewise
for the F$+-$.  Energies are given with respect to the S-wave
ground state (1S). Here $a=0.110(6)$~fm was used to convert the energies
to physical units. The error bars shown here contain statistical
errors from the lattice calculations and the uncertainty in the extracted
lattice spacing. The solid line %labelled as ``fit'' is from a model based
on the one-body Dirac equation -- see section \protect\ref{diracmodel} for details.}
\label{fig:espectr}
\end{figure}

\begin{table}
\centering
 \begin{tabular}{|c|ccc||c|ccc|}
 \hline
 nL$\pm -$1S &Lattice 1&Lattice 2&Lattice 3& nL$\pm -$1S  &Lattice 1&Lattice 2  &Lattice 3  \\
 \hline
 1S          & 0       & 0       & 0       & 2S           & 0.68(3) & 0.858(15) & 0.82(3)   \\
 1P$-$       & 0.46(4) & 0.42(2) & 0.43(3) & 2P$-$        & 1.40(6) & 1.38(2)   & 1.36(5)   \\
 1P$+$       & 0.61(4) & 0.50(3) & 0.42(4) & 2P$+$        & 1.60(5) & 1.44(3)   & 1.35(6)   \\
 1D$\pm$     & -       & 1.01(4) & 1.03(6) & 2D$\pm$      & -       & 1.54(2)   & 1.67(5)   \\
 1D$-$       & 0.93(6) & 0.86(3) & 0.86(5) & 2D$-$        & 1.83(4) & 1.771(15) & 1.79(4)   \\
 1D$+$       & 0.96(8) & 0.91(3) & 0.96(3) & 2D$+$        & 1.80(6) & 1.80(2)   & 1.85(2)   \\
 1F$\pm$     & 1.36(5) & 1.24(4) & 1.20(3) & 2F$\pm$      & 2.10(2) & 2.029(15) & 2.027(13) \\
 \hline
\end{tabular}
\caption{Heavy-light meson energy differences in GeV. Lattice 1 uses a static heavy quark,
whereas Lattices 2 and 3 are slightly smeared in the time direction. Lattice 2 uses APE
type smearing and Lattice 3 uses hypercubic blocking. See~\protect\cite{Latt2006} for
details. $a=0.110(6)$~fm was used to convert the energies to physical units. (Note: In
the first version of this article $a=0.115(3)$~fm was used.)}
\label{EnergyTable}
\end{table}

The energy spectrum obtained is shown in Fig.~\ref{fig:espectr} and in Table~\ref{EnergyTable}.
% Using different smearing for the heavy quark does not seem to change the energies
%too much - except for the P$+$ state.
There seems to be some spread in the extracted energy for the P$+$ state between the
different lattices, but the reason for this is not understood yet.
The energy of the D$+-$ state
was expected to be near the spin average of the D$-$ and D$+$ energies, but it turned out
to be a poor estimate of this average. Thus it is not clear whether or not the F$+-$ energy
is near the spin average of the two F-wave states, as was hoped. Our earlier results can
be found in Ref.~\cite{PRD69}.

One interesting point to note here is that the spin-orbit splitting of the P-wave states
is small, almost zero. We extracted the energy difference of the P$+$ and P$-$ states in
two different ways:
\begin{enumerate}
\item
\emph{Indirectly} by simply calculating the difference using the energies given by the fits
in Eq.~\ref{C2fit}, when the P$+$ and P$-$ data are fitted separately.

\item
Combining the P$+$ and P$-$ data and fitting the ratio
$\frac{\textrm{C}_2(\textrm{P}+)}{\textrm{C}_2(\textrm{P}-)}$, which enables us
to go \emph{directly} for the spin-orbit splitting, $m_\textrm{P$+$}-m_\textrm{P$-$}$.

\end{enumerate}
The results of the fits are $(-5 \pm 45)$~MeV and $(11 \pm 13)$~MeV, respectively.
D-wave spin-orbit splitting was also extracted in a similar manner, giving the results
$(102 \pm 54)$~MeV and $(124 \pm 18)$~MeV, respectively.

\begin{figure}
\centering
\includegraphics[angle=-90,width=0.57\textwidth]{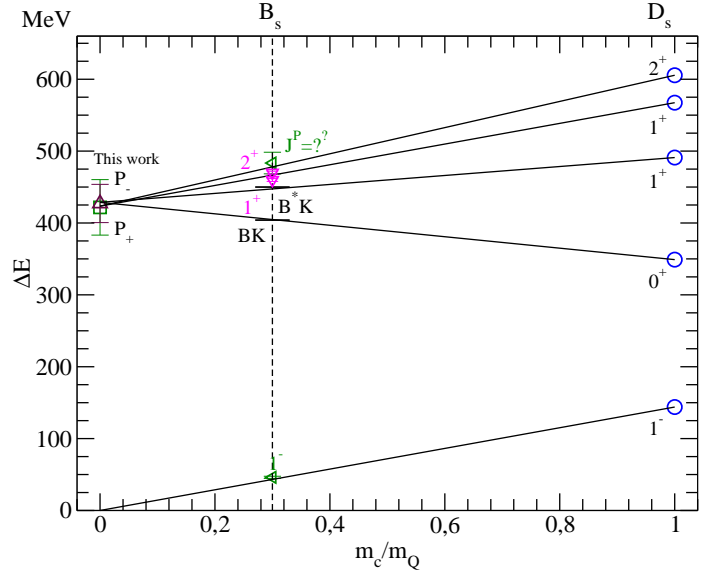}
\caption{Interpolation to the b-quark mass. The experimental data is from \protect\cite{PDG}.
For another lattice group's $m_c/m_Q=0$ results see~\protect\cite{II}.}
\label{fig:Bs_interpolation}
\end{figure}

To obtain predictions of the $B_s$ meson excited state energies, we can now interpolate between
the static heavy quark lattice calculations and $D_s$ meson experimental results, i.e. interpolate
between the static quark and the charm quark.  Linear
extrapolation seems to work well, as can be seen in Fig.~\ref{fig:Bs_interpolation}. The P-wave
states seem to lie near the $BK$ thresholds.

\section{Radial distributions: 3-point correlation function}

For evaluating the radial distributions of the light quark
a 3-point correlation function shown in Fig.~\ref{Fig:Corr} is needed. It is defined as
\begin{equation}
C_3(R,T)=\langle \Gamma^{\dag}\, U^Q\, \Gamma\, G_{q1}\, \Theta\,
G_{q2}\rangle.
\end{equation}
This is rather similar to the 2-point correlation function in Eq.~\ref{2point}.
We have now two light quark propagators, $G_{q1}$ and $G_{q2}$, and a
probe $\Theta(R)$ at distance $R$ from the static quark ($\gamma_4$ for the vector
(charge) and $1$ for the scalar (matter) distribution).

\begin{figure}[b]
\centering
 \includegraphics[angle=-90,width=0.495\textwidth]{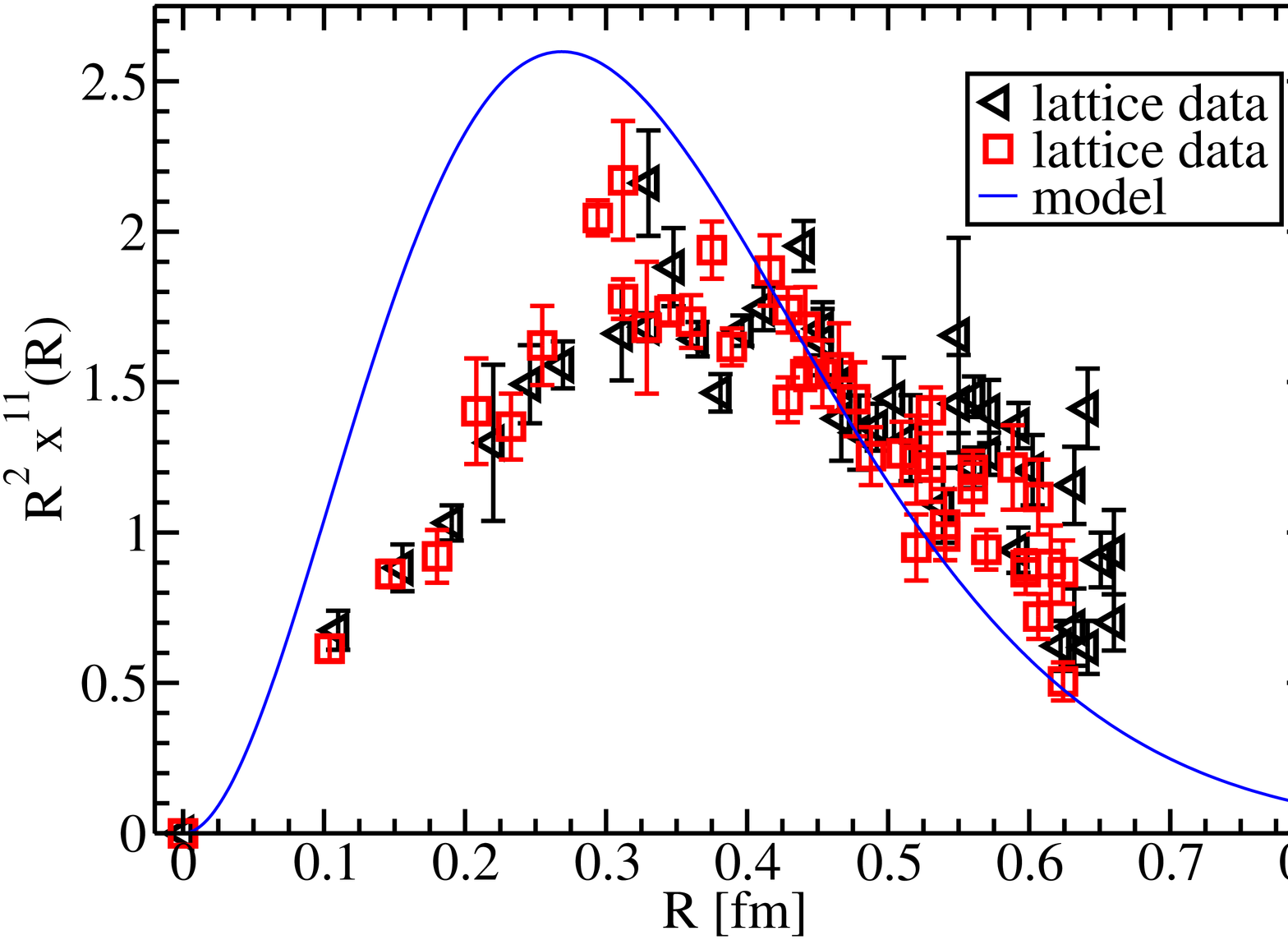}
 \includegraphics[angle=-90,width=0.495\textwidth]{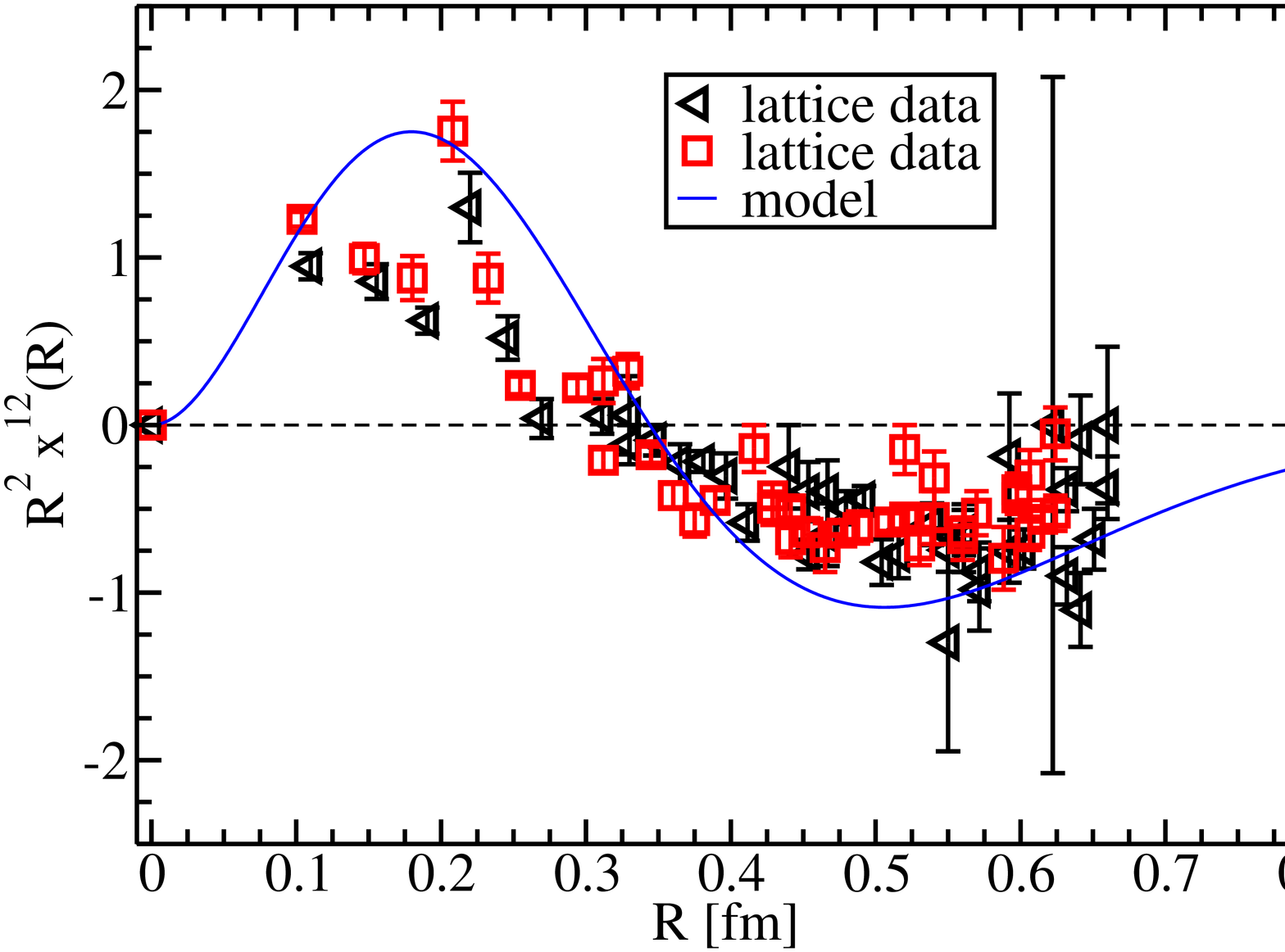}
\caption{On the left:
The S-wave ground state charge distribution. The label
``model'' on the solid line refers to the model presented in section~\protect\ref{diracmodel}.
On the right: The S-wave ground state and 1st excited state charge distribution overlap.
Note that we see one node, as expected from the Dirac equation.
}
\label{S11}
\end{figure}

\begin{figure}
\centering
\includegraphics[angle=-90,width=0.495\textwidth]{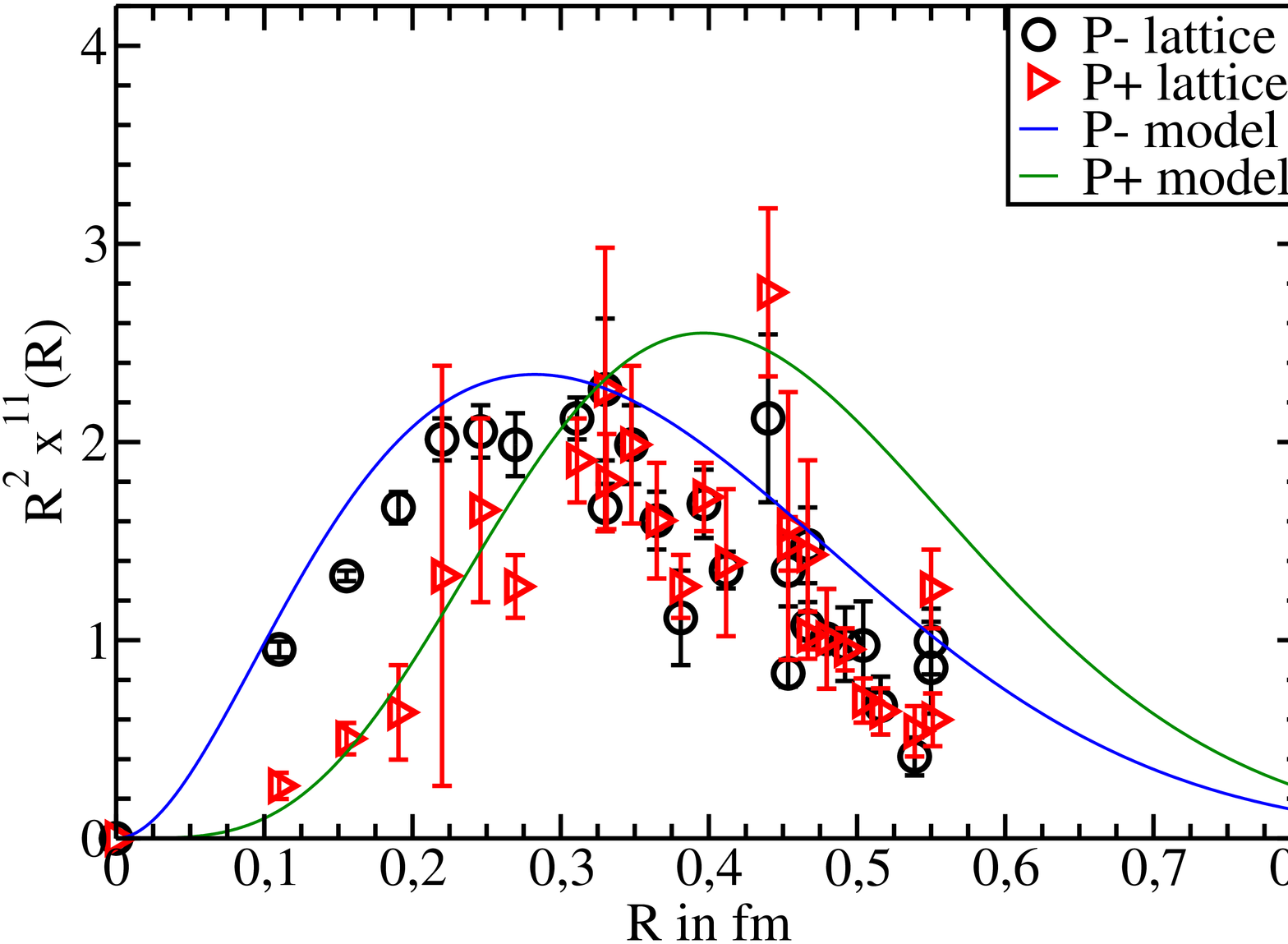}
\includegraphics[angle=-90,width=0.495\textwidth]{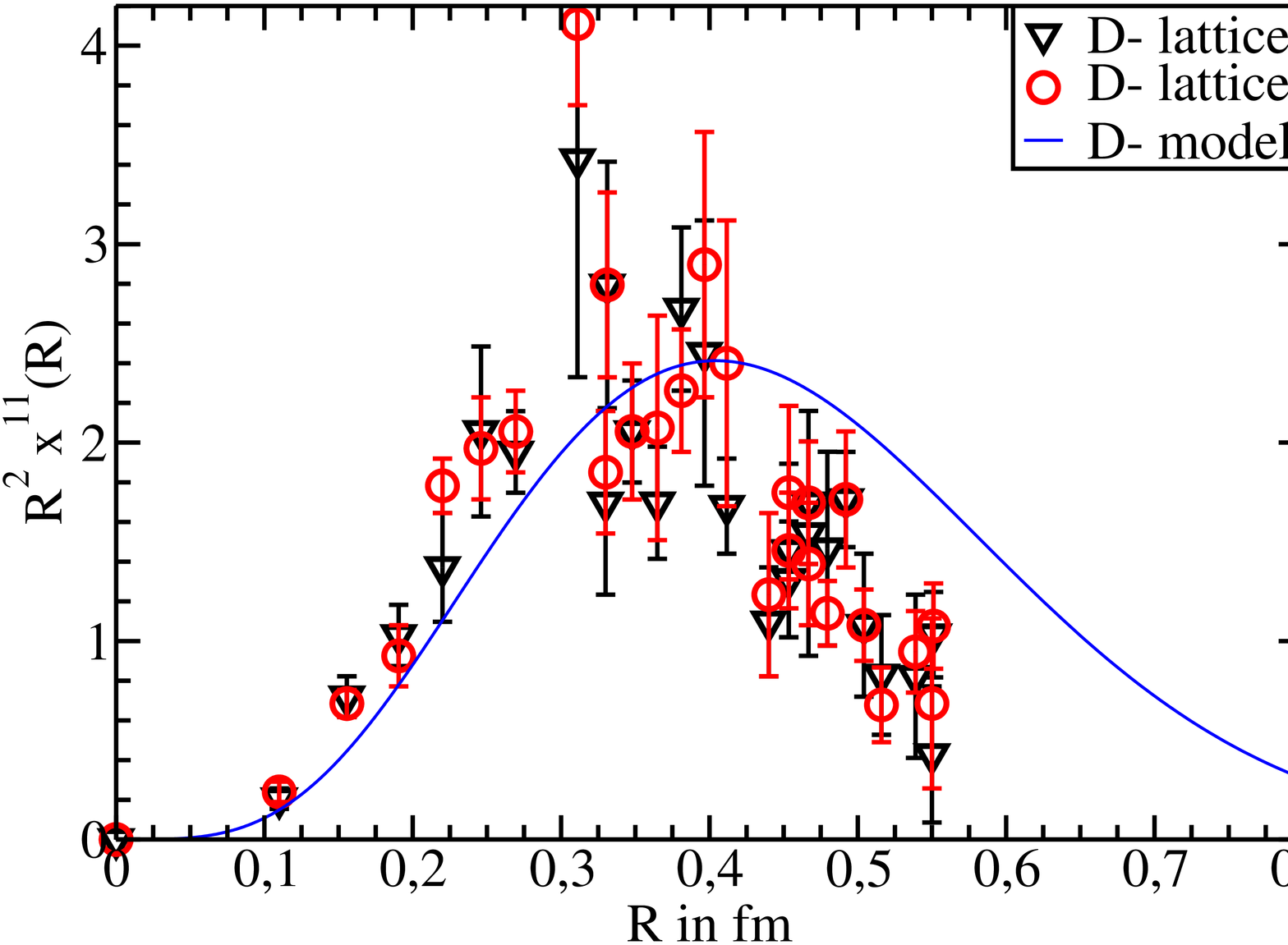}
\caption{On the left:
The P-wave ground state charge distributions. The P$+$ distribution has a peak
a bit further out than the P$-$, which is expected.
The solid lines are predictions from the model in section~\protect\ref{diracmodel}.
On the right:
The D$-$ ground state charge distribution.
% Smearing seems to slightly improve the measurements.
}
\label{PetD}
\end{figure}

Knowing the energies $m_i$ and the amplitudes $a_i$ from the earlier $C_2$ fit, 
the radial distributions, $x^{ij}(R)$'s, are then extracted by
fitting the $C_3$ with
\begin{equation}
C_3(R,T)\approx\sum_{i,j=1}^{N_{\textrm{max}}}a_{i}\mathrm{e}^{-m_i t_1}%
\; x^{ij}(R)\; \mathrm{e}^{-m_j t_2}a_{j}.
\end{equation}
The results are plotted in Figs.~\ref{S11} and~\ref{PetD}. The error bars in these
figures show statistical errors only.
See~\cite{PRD65etLatt} and \cite{Latt2006} for earlier calculations.

\section{A model based on the Dirac equation}
\label{diracmodel}

A simple model based on the Dirac equation is used to try
to describe the lattice data. Since the mass of the
heavy quark is infinite we have essentially a one-body
problem. The potential in the Dirac equation has a
linearly rising scalar part, $b_{\textrm{sc}} R$, as well as a
vector part $b_{\textrm{vec}} R$. The one gluon exchange potential,
$a_{\textrm{OGE}}\cdot V_{\textrm{OGE}}$, is modified to
\begin{equation}
V_{\textrm{OGE}}(R)\propto \int_0^{\infty}\!\! \mathrm{d}k\, j_0(kR)
\ln^{-1}\frac{k^2+4 m^2_g}{\Lambda^2_{\textrm{QCD}}},
\end{equation}
where $\Lambda_{\textrm{QCD}}=260$~MeV and
the dynamical gluon mass $m_g=290$~MeV (see~\cite{lahde} for details).
The potential also has a scalar term $m\omega L(L+1)$, which is needed
to increase the energy of higher angular momentum states. However, this
is only a small contribution (about 30~MeV for the F-wave).

The solid lines in the radial distribution plots are predictions from
the Dirac model fit with $m = 0.088$~GeV, $a_{\textrm{OGE}} = 0.81$,
$b_{\textrm{sc}} = 1.14$~GeV/fm, $b_{\textrm{vec}} \approx b_{\textrm{sc}}$
and $\omega = 0.028$. These are treated as free parameters with the values obtained
by fitting the ground state energies of P-, D- and F-wave states and the
energy of the first radially excited S-wave state (2S). Note that the
excited state energies in Fig.~\ref{fig:espectr} were not fitted.
%This fit was done using the energies obtained with APE smearing (``Sum6''), and
%the latest ``Hyp'' data was not used.

\section{Conclusions}

\begin{itemize}
\item
There is an abundance of lattice data, energies and radial distributions,
available for this $B_s$-like system.

\item
The P-wave spin-orbit splitting is small (essentially zero), which supports the symmetry
$b_{\textrm{vec}} = b_{\textrm{sc}}$ as proposed in~\cite{Page}. The D-wave spin-orbit
splitting is clearly non-zero and positive, whereas another lattice group seems to find
it to be slightly negative (see~\cite{II}).

\item
The energies and radial distributions of S-, P- and D-wave states can be
qualitatively understood by using a one-body Dirac equation.% model.
\end{itemize}

\section*{Acknowledgements}

I am grateful to my supervisor A. M. Green and to our collaborator,
Professor C. Michael.
I wish to thank the UKQCD Collaboration for providing the lattice
configurations.
I also wish to thank the Center for Scientific Computing in Espoo,
Finland, for making available the computer resources.
The EU grant HPRN-CT-2002-00311 Euridice
%and financial support from the Magnus Ehrnrooth foundation are
is also gratefully acknowledged.

\end{document}